\documentclass[aps,prapplied,reprint,groupedaddress,nofootinbib]{revtex4-2}

\usepackage{amsmath}
\usepackage{amssymb}
\usepackage{bm}
\usepackage{graphicx}
\usepackage{hyperref}
\usepackage{xcolor}
\hypersetup{hidelinks}

\newcommand{\ii}{\mathrm{i}}
\newcommand{\LO}{\mathrm{LO}}

\newcommand{\FSR}{\mathrm{FSR}}

\begin{document}

\title{Cavity-assisted homodyne detection with a single photodiode}

\author{Jonas Junker}
\affiliation{Institute for Applied Physics, Friedrich Schiller University Jena,
Albert-Einstein-Str.\ 15, 07745 Jena, Germany}
\affiliation{Abbe Center of Photonics, Friedrich Schiller University Jena,
Albert-Einstein-Str.\ 6, 07745 Jena, Germany}

\date{August 27, 2026}

\begin{abstract}
High-frequency squeezed states are important for quantum metrology and information processing, but quadrature measurements at gigahertz frequencies remain challenging. Balanced homodyne detection (BHD), the standard approach, requires closely matched complex transfer functions in both photodetection channels to suppress local-oscillator (LO) noise and maintain a well-defined readout quadrature. This matching becomes increasingly difficult at high bandwidths. Strongly asymmetric single-photodiode homodyne detection avoids this requirement, but achieving high signal efficiency and shot-noise clearance typically requires hundreds of milliwatts of LO power and does not suppress technical LO sidebands. I propose cavity-assisted homodyne detection, in which the quantum field and LO enter separate ports of an impedance-matched traveling-wave cavity. The resonant LO carrier is transmitted to the photodetector, where it beats with off-resonant signal sidebands reflected from the cavity. At the same time, technical LO sidebands outside the cavity linewidth are suppressed. I derive the quantum input--output relations and linearized photocurrent including intracavity loss, and show that near-unity signal-transfer efficiency can be achieved at moderate incident LO power. The cavity linewidth and free spectral range determine the usable detection band.
\end{abstract}

\maketitle

\section{Introduction}

Squeezed states of light have numerous applications, including quantum-enhanced heterodyne interferometry and frequency-multiplexed quantum information processing \cite{Gould2024,Eldan2026}. Accessing these states requires quadrature readout over analysis bands extending from megahertz to gigahertz frequencies.  Balanced homodyne detection (BHD) is the standard method for such measurements \cite{Lvovsky2009}.  To suppress common local-oscillator (LO) noise and maintain a well-defined readout quadrature across the analysis band, BHD requires closely matched complex transfer functions in its two photodetection channels.  Maintaining this matching becomes increasingly difficult at high frequencies.

Several approaches have pushed quadrature detection toward and into the multi-gigahertz regime, including integrated silicon photonics, phase-sensitive optical preamplification, and broadband balanced receivers \cite{Tasker2021,Inoue2023,Wang2025,Wang2026}. Direct squeezed-light measurements have reached 9 GHz with integrated homodyne detection and 43 GHz using optical preamplification \cite{Tasker2021,Inoue2023}. Achieving simultaneously high bandwidth and high quantum efficiency in conventional balanced detection, however, remains challenging. In a recent near-unity-quantum-efficiency detector, differential photodiode response limited balanced operation above approximately 500 MHz, whereas an unbalanced single-photodiode configuration provided shot-noise-limited detection up to 6.4 GHz \cite{Wilken2025}.

Single-photodiode homodyne detection therefore offers a direct way to avoid differential channel matching without optical preamplification.  It is also used in gravitational-wave dc readout, where an output mode cleaner transmits the measured field and a carrier component of that field serves as the LO \cite{Hild2009,Fricke2012}.  For separate quantum and LO fields, the simplest implementation combines them at a strongly asymmetric beam splitter and detects the signal-dominated output \cite{Wilken2024,Wilken2025}.  A highly asymmetric splitting ratio preserves most of the quantum signal but couples only a small fraction of the incident LO power to the detector.  Because the LO carrier and its technical noise sidebands experience the same attenuation, the relative technical LO noise is not suppressed.  A transmission mode cleaner can pre-filter the LO \cite{Willke1998}, but adds a separate resonator and leaves the fixed tradeoff between signal transfer and LO injection in place.

To separate signal transfer from LO injection spectrally, I propose the impedance-matched two-port traveling-wave cavity shown in Fig.~\ref{fig:setup}.  The quantum field and the LO enter through separate coupling mirrors.  The resonant LO carrier is transferred from the second port to the detector port.  Signal sidebands several cavity linewidths from resonance are reflected toward the same detector port with a signal efficiency approaching unity. Technical LO noise sidebands outside the cavity linewidth are suppressed.  The two fields interfere on a single photodiode, and their relative optical phase defines the measured quadrature.

The closest two-port geometry was demonstrated for mode-matched heterodyne characterization of a classical cavity, where a resonantly transmitted LO beats with a reflected frequency-offset probe \cite{Spector2024}.  Other cavity-based readouts, including optical ac coupling and self-homodyne or resonator-detection schemes, act on the carrier and sidebands of a single field \cite{Kwee2008,Kwee2009,Kaufer2017,Galatola1991,Zhang2000,Barbosa2013PRL, Barbosa2013PRA,Barbosa2020}.  Here, the cavity remains at a fixed operating point, resonant with an independently injected LO that interferes with a separate quantum field entering through the opposite port.

I derive the quantum input--output model and linearized photocurrent, including vacuum noise from intracavity loss. I then compare the cavity-assisted readout with an asymmetric beam splitter at equal detected LO power and signal efficiency. Because the cavity response is frequency selective, its free spectral range and linewidth set the accessible analysis band and achievable signal efficiency. The resulting design criteria identify regimes in which near-unity signal transfer, efficient LO injection, and suppression of technical LO noise can be achieved simultaneously.

\section{Measurement concept and field conventions}

\begin{figure}[t]
 \centering
 \includegraphics[width=\columnwidth]{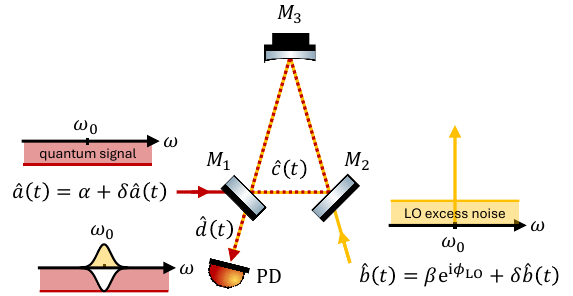}
 \caption{\label{fig:setup}Concept and schematic spectral response of cavity-assisted homodyne detection with a single photodiode. The quantum signal \(\hat a\) (red) enters through M1, while the LO \(\hat b\) (yellow) enters through M2. Both fields are mode matched to the same co-propagating cavity eigenmode, and M3 is nominally highly reflecting. The spectral sketches show the signal input (upper left), the LO carrier and its technical excess noise (right), and the detector-port field (lower left). At impedance matching, resonant signal components are suppressed at the M1 output, while the resonant LO carrier is transferred from M2 to the same port. Several cavity linewidths from resonance, signal sidebands are predominantly reflected, whereas technical LO amplitude-noise sidebands are suppressed.}
\end{figure}

The measurement architecture and its schematic spectral response are shown in Fig.~\ref{fig:setup}.  I consider a three-mirror traveling-wave ring cavity. Mirrors M1 and M2 are partially transmitting coupling mirrors, while M3 is nominally highly reflecting.  The quantum signal enters through M1 and the LO through M2.  Both fields are mode matched to the same co-propagating cavity eigenmode.  The outgoing field at M1 is detected with a single photodiode.

The two input fields are written as
\begin{align}
 \hat a(t)&=\alpha+\delta\hat a(t),\\
 \hat b(t)&=\beta e^{\ii\phi_{\LO}}+\delta\hat b(t).
\label{eq:input-fields}
\end{align}
The signal phase reference is chosen such that \(\alpha\) is real and nonnegative.  If no coherent signal carrier is present, this reference is set by the squeezing source or an auxiliary phase-reference field.  Here \(\alpha\geq0\) and \(\beta\geq0\) are the magnitudes of the coherent signal and LO amplitudes, respectively, \(\phi_{\LO}\) is the input LO phase relative to the signal reference, and \(\delta\hat a\) and \(\delta\hat b\) contain the corresponding sideband fluctuations.

\section{Cavity input--output model}

Near a selected longitudinal resonance, the relevant cavity mode obeys the standard input--output equation \cite{Gardiner1985}
\begin{equation}
 \dot{\hat c}=
 \left(\ii\Delta-\frac{\kappa}{2}\right)\hat c
 +\sqrt{\kappa_1}\hat a
 +\sqrt{\kappa_2}\hat b
 +\sqrt{\kappa_\ell}\hat v_\ell .
\label{eq:cavity-equation}
\end{equation}
Here
\begin{equation}
 \kappa=\kappa_1+\kappa_2+\kappa_\ell
\end{equation}
is the cavity intensity linewidth in angular-frequency units, \(\Delta=\omega_0-\omega_c\) is the carrier detuning, and \(\hat v_\ell\) is the vacuum field coupled through intracavity loss. The loss rate may be separated later into the small transmission of M3 and absorption or scattering, \(\kappa_\ell=\kappa_3+\kappa_{\mathrm{int}}\).

I use the Fourier convention
\begin{equation*}
 \hat a(t)=\int\frac{d\Omega}{2\pi}\,
 \hat a(\Omega)e^{-\ii\Omega t}.
\end{equation*}
Fourier transforming Eq.~\eqref{eq:cavity-equation} and solving algebraically for \(\hat c(\Omega)\) yields the cavity susceptibility
\begin{equation}
 \chi_c(\Omega)
 =\frac{1}{\kappa/2-\ii(\Omega+\Delta)}.
\label{eq:susceptibility}
\end{equation}
The field leaving at M1 is
\begin{equation}
 \hat d(\Omega)
 =r(\Omega)\hat a(\Omega)
 +t(\Omega)\hat b(\Omega)
 +l(\Omega)\hat v_\ell(\Omega),
\label{eq:output-field}
\end{equation}
with
\begin{align}
 r(\Omega)&=1-\kappa_1\chi_c(\Omega),\\
 t(\Omega)&=-\sqrt{\kappa_1\kappa_2}\,\chi_c(\Omega),\\
 l(\Omega)&=-\sqrt{\kappa_1\kappa_\ell}\,\chi_c(\Omega).
\label{eq:transfer-functions}
\end{align}
Thus, \(r\) describes reflection of the quantum signal at M1, \(t\) transfers the LO from M2 to the detector port at M1, and \(l\) couples intracavity-loss vacuum into that port. This phase convention follows from \(\hat d=\hat a-\sqrt{\kappa_1}\hat c\).  Any fixed propagation phase associated with injection through M2 can be incorporated into \(\hat b\).

The three coefficients satisfy
\begin{equation}
 |r(\Omega)|^2+|t(\Omega)|^2+|l(\Omega)|^2=1.
\label{eq:normalization}
\end{equation}
Equation~\eqref{eq:normalization} both expresses power conservation and preserves the commutator of the detected output field.  The loss-vacuum term must therefore be retained even when \(\kappa_\ell\) is small.

\subsection{Resonant and impedance-matched operation}

The general equations above allow arbitrary detuning.  I now set
\begin{equation}
 \Delta=0
\end{equation}
for the central measurement model.  Impedance matching at M1 requires
\begin{equation}
 r(0)=0
 \quad\Longleftrightarrow\quad
 \kappa_1=\kappa_2+\kappa_\ell .
\label{eq:matching}
\end{equation}
When internal loss is present, impedance matching does not require equal mirror transmissions.  Together with \(\kappa=\kappa_1+\kappa_2+\kappa_\ell\), Eq.~\eqref{eq:matching} gives \(\kappa=2\kappa_1\).  The coherent detector-port amplitude follows from the \(\Omega=0\) component of Eq.~\eqref{eq:output-field}.  Since the loss input has zero mean, \(\langle\hat v_\ell\rangle=0\),
\begin{equation*}
	\bar d\equiv\langle\hat d\rangle
	=r(0)\alpha+t(0)\beta e^{\ii\phi_{\LO}}.
\end{equation*}
At \(\Delta=0\), impedance matching gives \(r(0)=0\) and \(t(0)=-\sqrt{\kappa_2/\kappa_1}\), so
\begin{equation}
 \bar d
 =-\sqrt{\frac{\kappa_2}{\kappa_1}}\,
 \beta e^{\ii\phi_{\LO}} .
\label{eq:detected-carrier}
\end{equation}
With the signal phase reference chosen above, the detection phase is defined by \(\theta_{\mathrm{det}}=\arg\bar d\).  Under exact impedance matching, the input--output convention used here gives \(\theta_{\mathrm{det}}=\phi_{\LO}+\pi\).  This constant offset is absorbed in the experimental phase calibration. The resonant signal carrier proportional to \(\alpha\) is suppressed at the M1 output.  The resonant LO power transmission is
\begin{equation}
 T_{\LO}(0)\equiv|t(0)|^2=\frac{\kappa_2}{\kappa_1}
 =1-\frac{\kappa_\ell}{\kappa_1}.
\label{eq:lo-carrier-transmission}
\end{equation}
For \(\kappa_\ell\ll\kappa_1\), fixing the detected LO power requires only a small increase in incident LO power.

The resonant transfer functions reduce to
\begin{align}
 r(\Omega)&=\frac{-\ii\Omega}{\kappa/2-\ii\Omega},\\
 t(\Omega)&=
 -\frac{\sqrt{\kappa_1\kappa_2}}{\kappa/2-\ii\Omega},\\
 l(\Omega)&=
 -\frac{\sqrt{\kappa_1\kappa_\ell}}{\kappa/2-\ii\Omega}.
\label{eq:resonant-transfer}
\end{align}
For
\begin{equation}
 \kappa\ll|\Omega|\ll 2\pi\FSR ,
\label{eq:useful-band}
\end{equation}
where \(\FSR\) denotes the cavity free spectral range, the signal sidebands are promptly reflected, \(r(\Omega)\rightarrow1\), whereas \(t(\Omega),l(\Omega)\rightarrow0\).  The upper bound in Eq.~\eqref{eq:useful-band} excludes the neighboring longitudinal resonances.

\section{Photocurrent and measured quadrature}

I write the detected field as \(\hat d(t)=\bar d+\delta\hat d(t)\) and assume that its coherent amplitude is dominated by the transmitted LO, \(|t(0)|\beta\gg|r(0)|\alpha\).  This condition is automatic for an undisplaced input, \(\alpha=0\), and, for arbitrary \(\alpha\), under exact resonant impedance matching because \(r(0)=0\).  For a displaced input away from this operating point, the residual signal carrier must remain small compared with the detected LO carrier.  Direct photodetection gives a photocurrent proportional to \(\hat d^\dagger\hat d\).  In the usual strong-LO limit, retaining only terms linear in the field fluctuations gives
\begin{equation}
 \frac{\delta\hat I(\Omega)}{|\bar d|}
 =
 e^{-\ii\theta_{\mathrm{det}}}\delta\hat d(\Omega)
 +e^{\ii\theta_{\mathrm{det}}}\delta\hat d^\dagger(-\Omega).
\label{eq:linear-photocurrent}
\end{equation}
For a field \(\hat q\), the spectral quadrature at angle \(\theta\) is defined by
\begin{equation}
	\delta\hat X_{q,\theta}(\Omega)
	=
	e^{-\ii\theta}\delta\hat q(\Omega)
	+e^{\ii\theta}\delta\hat q^\dagger(-\Omega),
	\label{eq:spectral-quadrature}
\end{equation}
where vacuum noise is normalized to unity.

On resonance, \(r(-\Omega)=r^*(\Omega)\), and the same relation holds for \(t\) and \(l\).  Using the phase conventions above, Eq.~\eqref{eq:linear-photocurrent} becomes
\begin{equation}
 \begin{aligned}
 \frac{\delta\hat I(\Omega)}{|\bar d|}
 ={}&r(\Omega)
 \delta\hat X_{a,\theta_{\mathrm{det}}}(\Omega)\\
 &+t(\Omega)
 \delta\hat X_{b,\theta_{\mathrm{det}}}(\Omega)\\
 &+l(\Omega)
 \delta\hat X_{\ell,\theta_{\mathrm{det}}}(\Omega) .
 \end{aligned}
\label{eq:photocurrent-quadratures}
\end{equation}
Equation~\eqref{eq:photocurrent-quadratures} separates the detected photocurrent into contributions from the signal, LO, and internal-loss fields. All three are evaluated at the same detection quadrature \(\theta_{\mathrm{det}}\).  For the signal field, \(\theta_{\mathrm{det}}\) specifies the measured quadrature, whereas for the LO fluctuations it corresponds to the amplitude quadrature relative to the LO carrier.  The quadrature angle of the vacuum loss field is physically irrelevant.  The conjugate symmetry used above ensures that each transfer coefficient multiplies a spectral quadrature without mixing the orthogonal quadrature.

In the useful band of Eq.~\eqref{eq:useful-band},
\begin{equation}
	\frac{\delta\hat I(\Omega)}{|\bar d|}
	\longrightarrow
	\delta\hat X_{a,\theta_{\mathrm{det}}}(\Omega).
	\label{eq:homodyne-limit}
\end{equation}
The measurement therefore approaches homodyne readout of the signal quadrature at the selected detection phase.

For \(\Omega>0\), let \(S_I\) denote the one-sided photocurrent spectrum, normalized such that vacuum noise is unity, and let \(V_a(\Omega,\theta_{\mathrm{det}})\) denote the corresponding signal-quadrature spectrum.  For independent inputs, a coherent LO, and a vacuum loss port, Eq.~\eqref{eq:photocurrent-quadratures} gives
\begin{align}
	S_I^{\mathrm{opt}}(\Omega,\theta_{\mathrm{det}})
	&=|r|^2V_a+|t|^2+|l|^2 \notag\\
	&=1+\eta_{\mathrm{cav}}(\Omega)
	\left[V_a(\Omega,\theta_{\mathrm{det}})-1\right].
	\label{eq:measured-spectrum}
\end{align}
Here \(\eta_{\mathrm{cav}}(\Omega)\equiv|r(\Omega)|^2\).  For a squeezed state with squeezing angle \(\theta_s(\Omega)\), defined relative to the chosen signal phase reference,
\begin{equation}
	\begin{aligned}
		V_a(\Omega,\theta_{\mathrm{det}})
		&=V_-(\Omega)\cos^2[\theta_{\mathrm{det}}-\theta_s(\Omega)]\\
		&\quad+V_+(\Omega)\sin^2[\theta_{\mathrm{det}}-\theta_s(\Omega)].
	\end{aligned}
	\label{eq:squeezed-spectrum}
\end{equation}
The normalized photocurrent in Eq.~\eqref{eq:photocurrent-quadratures} is a linear field observable whose form is independent of the input state.  Its variance, however, depends on both the state and the detection phase, as shown in Eq.~\eqref{eq:squeezed-spectrum}.  For vacuum input, Eq.~\eqref{eq:normalization} gives \(S_I^{\mathrm{opt}}=1\) at all frequencies.

Let \(E_{\LO}\) denote the technical LO amplitude excess-noise PSD, normalized to the shot-noise PSD at the same detected LO power, and let \(S_{\mathrm{el}}\) denote the electronic noise in the same units.  The total measured spectrum is
\begin{equation}
	S_I^{\mathrm{tot}}
	=S_I^{\mathrm{opt}}
	+|H_{\LO}(\Omega)|^2E_{\LO}(\Omega)
	+S_{\mathrm{el}}(\Omega),
	\label{eq:total-spectrum}
\end{equation}
where \(H_{\LO}(\Omega)=t(\Omega)/t(0)\) is the LO-sideband transfer function normalized to the transmitted carrier and defined in Eq.~\eqref{eq:lo-filter} below.

I use one-sided spectra throughout.  If \(S_{\mathrm{RIN}}(\Omega)\) is the one-sided technical relative-intensity-noise PSD of the LO, with the shot-noise contribution excluded, then
\begin{equation}
	E_{\LO}(\Omega)=\frac{I_{\mathrm{det}}S_{\mathrm{RIN}}(\Omega)}{2e},
	\qquad
	I_{\mathrm{det}}=\mathcal R P_{\LO,\mathrm{det}},
	\label{eq:rin-normalization}
\end{equation}
for a linear photodiode of responsivity \(\mathcal R\).  Here \(2eI_{\mathrm{det}}\) is the corresponding one-sided photocurrent shot-noise PSD.  Thus, \(E_{\LO}\) contains only technical excess noise.  The vacuum contribution is already included in Eq.~\eqref{eq:measured-spectrum}. Because \(I_{\mathrm{det}}\) is the actual dc photocurrent, Eq.~\eqref{eq:rin-normalization} already includes the photodiode responsivity and quantum efficiency.  A detector-efficiency factor introduced later in the signal transfer must therefore not be applied again to \(E_{\LO}\).

For \(|\Omega|\gg\kappa\), the LO-input sidebands no longer contribute to the normalized photocurrent because \(t(\Omega)\rightarrow0\).  For a vacuum signal input, the unit optical vacuum noise is then supplied predominantly by the signal port, while the detected LO carrier continues to set the homodyne gain and hence the absolute photocurrent-noise level.

\section{Signal transfer and local-oscillator filtering}

\begin{figure}[t]
	\centering
	\includegraphics[width=\columnwidth]{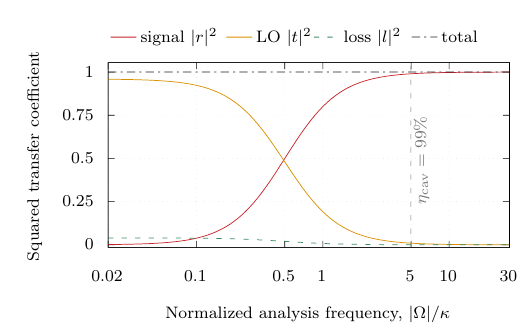}
	\caption{\label{fig:port-vacuum}
		Squared detector-port transfer coefficients for a resonant, impedance-matched
		cavity with \(\kappa_1/\kappa=0.50\), \(\kappa_2/\kappa=0.48\), and
		\(\kappa_\ell/\kappa=0.02\).  The red curve is the signal-port coefficient
		\(\eta_{\mathrm{cav}}=|r|^2\).  The orange curve is the LO-port coefficient
		\(|t|^2=T_{\LO}(0)|H_{\LO}|^2\), and the green curve is the internal-loss
		coefficient \(|l|^2\).  For vacuum inputs, these coefficients partition the
		unit detector-port vacuum-noise level.  The black dash-dotted line shows
		their sum.  The vertical gray dotted line marks
		\(\eta_{\mathrm{cav}}=0.99\).}
\end{figure}

For the resonant, impedance-matched case considered above, the cavity signal-transfer efficiency is
\begin{equation}
	\eta_{\mathrm{cav}}(\Omega)
	=|r(\Omega)|^2
	=\frac{4(\Omega/\kappa)^2}
	{1+4(\Omega/\kappa)^2}.
	\label{eq:signal-efficiency}
\end{equation}
To quantify technical LO noise relative to the transmitted carrier, I define the normalized LO-sideband transfer function as
\begin{equation}
	\begin{aligned}
		H_{\LO}(\Omega)&\equiv\frac{t(\Omega)}{t(0)},\\
		|H_{\LO}(\Omega)|^2&=\frac{1}{1+4(\Omega/\kappa)^2}.
	\end{aligned}
	\label{eq:lo-filter}
\end{equation}
The signal efficiency and normalized LO-sideband transfer therefore satisfy
\begin{equation}
	\eta_{\mathrm{cav}}(\Omega)+|H_{\LO}(\Omega)|^2=1.
	\label{eq:complementarity}
\end{equation}
The quantity \(|H_{\LO}|^2\) is normalized to the transmitted carrier.  The actual LO-port contribution to the detected vacuum noise is \[ |t(\Omega)|^2 =T_{\LO}(0)|H_{\LO}(\Omega)|^2. \] Internal loss does not alter the normalized relation in Eq.~\eqref{eq:complementarity} under exact impedance matching, but reduces the resonant LO transmission according to Eq.~\eqref{eq:lo-carrier-transmission}.  From \(l(\Omega)\) in Eq.~\eqref{eq:transfer-functions}, the contribution of the internal-loss port to the detected vacuum noise is
\begin{equation}
	|l(\Omega)|^2
	=
	\frac{\kappa_1\kappa_\ell}
	{\Omega^2+(\kappa/2)^2}.
	\label{eq:loss-vacuum}
\end{equation}
The resulting squared detector-port transfer coefficients are shown in Fig.~\ref{fig:port-vacuum}.

For vacuum inputs, the signal-port contribution vanishes on resonance, and the LO and internal-loss ports together supply the unit vacuum-noise level.  As the analysis frequency moves several linewidths from the selected resonance, the signal-port contribution approaches unity while the LO- and internal-loss-port contributions become small.  At the vertical line, \(\eta_{\mathrm{cav}}=0.99\), and Eq.~\eqref{eq:complementarity} gives \(|H_{\LO}|^2=0.01\).  Thus, the same operating point that preserves \(99\%\) of the signal-sideband power suppresses technical LO-noise power by \(20\,\mathrm{dB}\) relative to the transmitted carrier.

\section{Comparison with an asymmetric beam splitter}

For a lossless beam splitter, let \(R\) denote the signal efficiency, i.e., the fraction of incident signal power directed to the detected output.  The corresponding field is
\begin{equation}
	\hat d_{\mathrm{BS}}
	=
	\sqrt{R}\,\hat a
	+\sqrt{1-R}\,\hat b ,
	\label{eq:bs-output}
\end{equation}
where a fixed beam-splitter phase has again been absorbed into the field definitions.  In the strong-LO limit,
\begin{equation}
	\frac{\delta\hat I_{\mathrm{BS}}}{|\bar d_{\mathrm{BS}}|}
	=
	\sqrt{R}\,\delta\hat X_{a,\theta_{\mathrm{det}}}
	+\sqrt{1-R}\,\delta\hat X_{b,\theta_{\mathrm{det}}}.
	\label{eq:bs-measurement}
\end{equation}
Because the LO carrier and its technical amplitude-noise sidebands are multiplied by the same beam-splitter coefficient, their relative transfer is
\begin{equation}
	H_{\LO,\mathrm{BS}}(\Omega)=1.
	\label{eq:bs-lo-transfer}
\end{equation}

I compare the two readouts at equal detected LO power, using the same photodiode and electronic readout and assuming equal net spatial-mode overlap.  This gives the same homodyne gain for a given field fluctuation at the detector port.  When the detected carrier is LO-dominated, equal detected LO power also implies equal shot-noise level, mean photocurrent, and detector loading.  Equal gain from the incident signal field additionally requires equal signal efficiency, as imposed below.  For resonant cavity operation, the incident LO powers required to obtain the same detected LO power are
\begin{align}
	\frac{P_{\LO,\mathrm{in}}^{\mathrm{BS}}}
	{P_{\LO,\mathrm{det}}}
	&=\frac{1}{1-R},\\
	\frac{P_{\LO,\mathrm{in}}^{\mathrm{cav}}}
	{P_{\LO,\mathrm{det}}}
	&=\frac{1}{|t(0)|^2}
	=\frac{\kappa^2}{4\kappa_1\kappa_2}.
	\label{eq:incident-lo-power}
\end{align}
For the asymmetric beam splitter, the incident-to-detected LO power ratio is \(20\), \(50\), and \(100\) for signal efficiencies of 95\%, 98\%, and 99\%, respectively.  Under exact impedance matching, the corresponding ratio for the cavity-assisted readout reduces to \(\kappa_1/\kappa_2\) and differs from unity only because of intracavity loss.  Here \(P_{\LO,\mathrm{in}}^{\mathrm{cav}}\) denotes the LO power in the spatial mode matched to the cavity through M2.  External mode mismatch would increase the required incident power.

The cavity reaches the same signal efficiency \(R\) when
\begin{equation}
	\frac{|\Omega|}{\kappa}
	=
	\frac{1}{2}\sqrt{\frac{R}{1-R}}.
	\label{eq:equal-efficiency}
\end{equation}
At the same analysis frequency,
\begin{equation}
	|H_{\LO}(\Omega)|^2=1-R.
	\label{eq:equal-efficiency-filter}
\end{equation}
For \(R=0.99\), Eq.~\eqref{eq:equal-efficiency} gives \(|\Omega|/\kappa\simeq4.97\).  At this frequency, both readouts have a 99\% signal efficiency, while the cavity suppresses technical LO amplitude-noise power by a factor of \(100\), or \(20\,\mathrm{dB}\), relative to the unfiltered beam-splitter readout.

To compare the two readouts independently of the measured quantum state, I refer all added noise to the signal input.  Neglecting common detector factors for clarity, I define
\begin{equation}
	S_{I,\mathrm{in}}^{\mathrm{eq}}
	=\frac{S_I^{\mathrm{tot}}}{\eta}
	=V_a+N_{\mathrm{add}},
	\label{eq:input-referred-definition}
\end{equation}
so that \(N_{\mathrm{add}}\) includes the vacuum noise introduced by incomplete signal transfer.  For both readouts, let \(E_{\LO}\) denote the technical LO amplitude excess-noise PSD of the unfiltered LO, referred to the detector and normalized to the shot-noise PSD at the chosen detected LO power.  Because the beam splitter attenuates the LO carrier and its technical sidebands by the same factor, this normalized excess noise is unchanged by the beam splitter.  Let \(S_{\mathrm{el}}\) denote the electronic noise in the same units.  The two readouts then give
\begin{align}
	N_{\mathrm{add}}^{\mathrm{cav}}
	&=
	\frac{1-\eta_{\mathrm{cav}}
		+|H_{\LO}|^2E_{\LO}+S_{\mathrm{el}}}
	{\eta_{\mathrm{cav}}},
	\label{eq:added-noise-cavity}\\
	N_{\mathrm{add}}^{\mathrm{BS}}
	&=
	\frac{1-R+E_{\LO}+S_{\mathrm{el}}}{R}.
	\label{eq:added-noise-bs}
\end{align}
At equal signal efficiency, the vacuum and electronic contributions are the same, whereas the technical LO amplitude-noise contribution of the cavity-assisted readout is smaller by the factor \(1-R\).

The beam-splitter readout can also use an LO pre-filtered by a transmission mode cleaner.  For a direct comparison, I consider a second cavity with the same parameters as the cavity-assisted readout, including the linewidth and free spectral range.  It multiplies \(E_{\LO}\) by the same normalized LO-sideband transfer while leaving the signal path unchanged \cite{Willke1998}.  At the equal-efficiency point, this filtered beam-splitter readout provides the same technical-noise suppression as the cavity-assisted readout while retaining the frequency-independent signal efficiency \(R\).  If \(T_{\mathrm{MC}}\) denotes the carrier transmission of the mode cleaner, the required LO power upstream of it exceeds the detected LO power by the factor
\begin{equation}
	\frac{P_{\LO,\mathrm{in}}^{\mathrm{MC+BS}}}
	{P_{\LO,\mathrm{det}}}
	=
	\frac{1}{T_{\mathrm{MC}}(1-R)}.
	\label{eq:mode-cleaner-lo-power}
\end{equation}
By comparison, the cavity-assisted readout performs the filtering and field combination in a single resonator and transfers, rather than discards, the resonant LO carrier.  It is also not limited by a fixed splitting ratio. Beyond the equal-efficiency point, the cavity signal efficiency exceeds \(R\), so its vacuum penalty \(1-\eta_{\mathrm{cav}}\) falls below the fixed beam-splitter penalty \(1-R\).  At frequencies well above the filter linewidth and away from neighboring cavity resonances, the technical LO contribution vanishes for either filtered scheme, while the beam-splitter readout approaches
\begin{equation*}
	N_{\mathrm{add}}
	=
	\frac{1-R+S_{\mathrm{el}}}{R}.
\end{equation*}
Table~\ref{tab:squeezing-budget} quantifies this remaining difference.

\section{Practical comparison}

To evaluate the general relations above for a finite-length cavity with internal loss, consider a triangular ring cavity operated at \(1064\,\mathrm{nm}\) with round-trip length \(L_{\mathrm{rt}}=0.20\,\mathrm{m}\), corresponding to \(\FSR=1.499\,\mathrm{GHz}\).  A finesse of \(\mathcal F=1000\) gives a power linewidth of \(1.50\,\mathrm{MHz}\).  This linewidth is a representative design choice rather than an optimum.  For fixed FSR, a narrower linewidth would move the high-efficiency band toward lower frequencies, but would generally also tighten the detuning tolerance, increase the circulating power needed for a fixed detected LO power, and make fixed intracavity loss more important, as discussed in Sec.~VIII.  A simple near-matched realization uses equal coupling-mirror power transmissions \(T_1=T_2\simeq0.304\%\), corresponding to power reflectivities of approximately \(99.696\%\), and a round-trip internal power loss \(\mathcal L_{\mathrm{int}}=200\,\mathrm{ppm}\).  Because of the finite loss, the signal port is slightly undercoupled rather than exactly impedance matched: the resonant LO transmission is \(T_{\LO}(0)\simeq0.937\), while the residual resonant signal reflection is \(|r(0)|^2\simeq1.0\times10^{-3}\).

\begin{figure}[t]
	\includegraphics[width=\columnwidth]{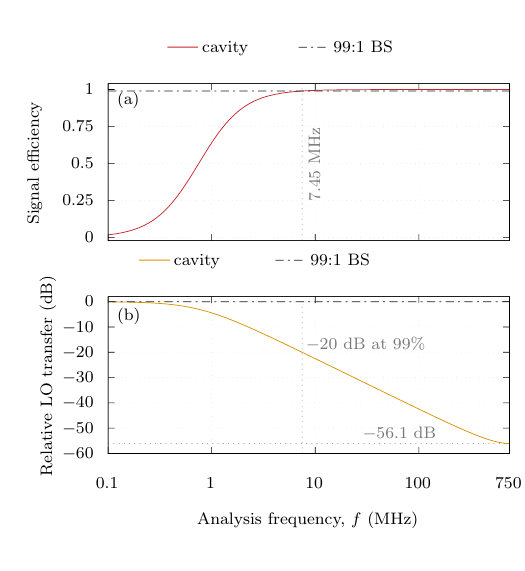}
	\caption{\label{fig:representative-transfer}
		Frequency-dependent response of the practical equal-coupler cavity from
		resonance to the first antiresonance, with
		\(L_{\mathrm{rt}}=0.20\,\mathrm m\), \(\mathcal F=1000\),
		\(T_1=T_2\simeq0.304\%\), and round-trip internal power loss
		\(\mathcal L_{\mathrm{int}}=200\,\mathrm{ppm}\).
		(a) Signal efficiency compared with a 99:1 beam splitter, which directs
		99\% of the incident quantum-signal power to the detected output.  The vertical
		line marks the first frequency at which the cavity reaches the same
		99\% signal efficiency, \(f=7.45\,\mathrm{MHz}\).
		(b) Technical LO amplitude-noise transfer, normalized to the transmitted LO
		carrier, compared with the unfiltered 99:1 beam-splitter readout.  At the
		99\% signal-efficiency point, the cavity suppresses the relative LO
		amplitude-noise power by \(20\,\mathrm{dB}\).  At the antiresonance, the
		suppression reaches \(56.1\,\mathrm{dB}\).  The curves are calculated using
		the periodic cavity response of Eqs.~\eqref{eq:periodic-lo-transfer} and
		\eqref{eq:periodic-equal-coupler-reflection}.  Detector and electronic noise
		are not included.}
\end{figure}

Figure~\ref{fig:representative-transfer} shows the periodic cavity response from resonance to the first antiresonance at \(f=\FSR/2=749.5\,\mathrm{MHz}\).  Over this interval, the signal efficiency increases monotonically and the technical LO-noise transfer decreases monotonically.  At \(f=7.45\,\mathrm{MHz}\), the cavity-assisted readout reaches the 99\% signal efficiency of the beam-splitter readout and suppresses its technical LO amplitude-noise power by \(20\,\mathrm{dB}\).  At the antiresonance, the periodic model of Appendix~\ref{app:detuning} gives \(\eta_{\mathrm{cav}}=0.9999975\) and \(|H_{\LO}^{\mathrm{per}}|^2=2.47\times10^{-6}\), corresponding to \(56.1\,\mathrm{dB}\) of suppression.  The incident-to-detected LO power ratio is \(1.067\), compared with 100 for the 99:1 beam splitter.  Because the response is symmetric about the antiresonance and repeats at each FSR, the signal efficiency remains at least 99\% from \(7.45\,\mathrm{MHz}\) to \(\FSR-7.45\,\mathrm{MHz}\simeq1.49\,\mathrm{GHz}\) within the first free spectral range.  The cavity therefore provides broad high-efficiency bands separated by narrow regions around successive resonances, rather than a continuous broadband response over multiple FSRs.

\begin{figure}[t]
	\includegraphics[width=\columnwidth]{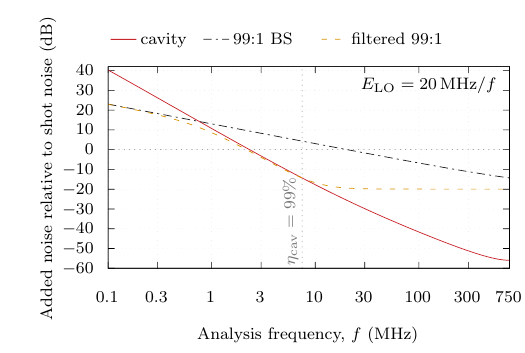}
	\caption{\label{fig:total-noise-comparison}
		Input-referred added noise from
		Eqs.~\eqref{eq:added-noise-cavity} and \eqref{eq:added-noise-bs} for the
		practical cavity of Fig.~\ref{fig:representative-transfer}, a 99:1 beam
		splitter, and the same beam splitter with the LO pre-filtered by an identical
		second cavity, all at equal detected LO power.  The
		filtered readout floors at \((1-R)/R=-20\,\mathrm{dB}\).  The plotted quantity is
		\(N_{\mathrm{add}}\), not the total noise \(V_a+N_{\mathrm{add}}\).  The
		calculation uses \(P_{\LO,\mathrm{det}}=5\,\mathrm{mW}\),
		\(E_{\LO}(f)=20\,\mathrm{MHz}/f\), and \(S_{\mathrm{el}}=0\).  The vertical
		line marks \(\eta_{\mathrm{cav}}=0.99\), where both schemes have the same
		signal efficiency.  The \(1/f\) spectrum is an illustrative technical
		LO-noise model.  Below approximately \(2\,\mathrm{MHz}\),
		relaxation-oscillation and control-loop features are laser specific.  The
		plotted interval ends at the cavity antiresonance, \(f=\FSR/2\).}
\end{figure}

For the added-noise comparison, I model the normalized technical LO excess-noise spectrum as \(E_{\LO}(f)=20\,\mathrm{MHz}/f\) at a detected LO power of \(5\,\mathrm{mW}\).  Thus, the technical excess noise equals the shot-noise PSD at \(20\,\mathrm{MHz}\).  At \(1064\,\mathrm{nm}\), this optical power gives a mean photodiode current \(I_{\mathrm{det}}\simeq4.3\,\mathrm{mA}\) for unit quantum efficiency.  The chosen crossover frequency is consistent with measurements of a diode-pumped single-frequency Nd:YAG laser that reached shot-noise-limited relative intensity noise above \(20\,\mathrm{MHz}\) at a photocurrent of \(4.4\,\mathrm{mA}\)~\cite{Kane1990}.  The \(1/f\) model is illustrative rather than a fit.  Measured NPRO spectra exhibit laser- and stabilization-dependent structure at lower frequencies \cite{Vahlbruch2025NPRO}. Figure~\ref{fig:total-noise-comparison} evaluates Eqs.~\eqref{eq:added-noise-cavity} and \eqref{eq:added-noise-bs} for the practical cavity of Fig.~\ref{fig:representative-transfer} and a 99:1 beam splitter, with common detector factors omitted and \(S_{\mathrm{el}}=0\). At the equal-efficiency point, \(f=7.45\,\mathrm{MHz}\), the input-referred added noise is \(-14.3\,\mathrm{dB}\) for the cavity-assisted readout and \(+4.3\,\mathrm{dB}\) for the unfiltered 99:1 beam-splitter readout, both relative to shot noise.  The comparison also identifies the low-frequency region in which the signal-extraction penalty of the cavity-assisted readout makes the beam-splitter readout preferable.

For the same parameters, consider a 99:1 readout whose LO is pre-filtered by an identical second cavity operated as a transmission mode cleaner.  Its technical LO excess-noise spectrum is \(E_{\LO}|H_{\LO}^{\mathrm{per}}|^2\), with the periodic transfer function of Eq.~\eqref{eq:periodic-lo-transfer}.  Within this model, the resulting added noise equals that of the cavity-assisted readout at the equal-efficiency point. At higher frequencies, it approaches \((1-R)/R=-20\,\mathrm{dB}\), while the added noise of the cavity-assisted readout continues to fall as its signal efficiency increases. With the detector-referred normalization of Eq.~\eqref{eq:rin-normalization}, the detector-level variance used in Table~\ref{tab:squeezing-budget} is \[ V_{\mathrm{det}} =1+\eta_{\mathrm{pd}}\eta_{\mathrm{mm}}\eta(V_a-1) +|H|^2E_{\LO}. \] Here \(\eta=\eta_{\mathrm{cav}}\) for the cavity-assisted readout and \(\eta=R\) for either beam-splitter readout.  The transfer is \(H=H_{\LO}^{\mathrm{per}}\) for both filtered schemes and \(H=1\) for the unfiltered beam splitter.  Thus, \(\eta_{\mathrm{pd}}\) is not applied again to the detector-referred technical LO term. Table~\ref{tab:squeezing-budget} gives the calculated detected quadrature-noise levels for a \(10\,\mathrm{dB}\) squeezed input at three frequencies, with and without the stated photodiode loss.  The residual advantage over the filtered beam-splitter readout arises from the vacuum penalties \(1-\eta_{\mathrm{cav}}\) and \(1-R\), and is appreciable only when the common detection efficiency is close to unity.  By contrast, the advantage over the unfiltered beam-splitter readout remains substantial wherever technical LO excess noise is non-negligible.

\begin{table}[t]
	\caption{\label{tab:squeezing-budget}
		Calculated detected quadrature-noise level in dB relative to shot noise for
		a \(10\,\mathrm{dB}\) squeezed input (\(V_-=0.10\)), with
		\(E_{\LO}=20\,\mathrm{MHz}/f\),
		\(S_{\mathrm{el}}=0\), and phase noise excluded, for the practical cavity of
		Fig.~\ref{fig:representative-transfer}, a 99:1 beam splitter with the LO
		pre-filtered by an identical second cavity, and the
		unfiltered 99:1 beam splitter.  Values assume a photodiode efficiency
		\(\eta_{\mathrm{pd}}=0.97\) and unit signal--LO modal overlap
		\(\eta_{\mathrm{mm}}=1\).  Values in parentheses set
		\(\eta_{\mathrm{pd}}=1\) while keeping the detector-referred
		\(E_{\LO}\) fixed.  Positive values indicate noise above shot noise.}
	\begin{ruledtabular}
		\begin{tabular}{lccc}
			\(f\) & Cavity & Filtered 99:1 & Unfiltered 99:1 \\
			\hline
			\(10\,\mathrm{MHz}\)    & \(-8.4\) \((-9.3)\)  & \(-8.3\) \((-9.2)\) & \(+3.3\) \((+3.2)\) \\
			\(100\,\mathrm{MHz}\)   & \(-9.0\) \((-10.0)\) & \(-8.7\) \((-9.6)\) & \(-4.7\) \((-5.1)\) \\
			\(749.5\,\mathrm{MHz}\) & \(-9.0\) \((-10.0)\) & \(-8.7\) \((-9.6)\) & \(-7.9\) \((-8.7)\)
		\end{tabular}
	\end{ruledtabular}
\end{table}

\section{Practical considerations and limitations}

\paragraph{Impedance mismatch.}
Let
\begin{equation}
	\mu=\kappa_1-(\kappa_2+\kappa_\ell).
\end{equation}
At zero detuning,
\begin{equation}
	|r(\Omega)|^2
	=
	\frac{\Omega^2+\mu^2/4}
	{\Omega^2+\kappa^2/4}.
	\label{eq:mismatch}
\end{equation}
For \(\kappa\ll|\Omega|\) within the single-resonance approximation, the reflection still approaches unity.  A small mismatch has little effect on the reflected sidebands, but it leaves a residual carrier from the quantum-signal input at the detector port, with \(|r(0)|^2=\mu^2/\kappa^2\).  This residual carrier matters primarily for a bright or displaced input and is absent for an undisplaced field.

\paragraph{Detuning and cavity-length noise.}
A constant detuning shifts the filter response and detected optical phase. For \(\Delta\ne0\), \(r(-\Omega)\ne r^*(\Omega)\), which produces frequency-dependent signal-quadrature mixing.  This effect is negligible when both sidebands remain far from the shifted resonance and must otherwise be included in the calibrated transfer function.

The LO sidebands also lose conjugate symmetry, allowing LO phase noise to be converted into detected amplitude noise.  A fluctuating detuning additionally changes the detected LO phase close to resonance by
\begin{equation}
	\delta\theta_{\mathrm{det}}
	\simeq\frac{2\,\delta\Delta}{\kappa},
	\label{eq:detuning-phase}
\end{equation}
up to the sign convention for \(\Delta\).  A narrower cavity gives stronger LO-sideband rejection but greater sensitivity to length fluctuations.  For operation at the squeezed quadrature and a small, zero-mean rms detection-phase error \(\sigma_\theta\), Eq.~\eqref{eq:squeezed-spectrum} gives
\begin{equation}
	\overline{V}_a
	\simeq V_-+\sigma_\theta^2\left(V_+-V_-\right),
	\qquad
	\sigma_\theta\simeq\frac{2\sigma_\Delta}{\kappa},
	\label{eq:phase-noise-mixing}
\end{equation}
so that even weak phase noise can mix anti-squeezing into the measured quadrature.  Equation~\eqref{eq:phase-noise-mixing} gives the corresponding phase-stability requirement when the linewidth is reduced.

For illustration, a state with \(V_-=0.10\) (\(10\,\mathrm{dB}\) squeezing) and \(V_+=31.6\) (\(15\,\mathrm{dB}\) anti-squeezing) permits \(\sigma_\theta\lesssim1.1^\circ\), or \(\sigma_\Delta/\kappa\lesssim0.01\), for less than \(0.5\,\mathrm{dB}\) of phase-noise degradation in this quasi-static estimate.  For the design of Sec.~VII, this tolerance corresponds to an rms detuning of \(15\,\mathrm{kHz}\), i.e., a required rms round-trip length stability of about \(11\,\mathrm{pm}\).  Non-common path-length fluctuations between the signal and LO injection paths add to the detection-phase error with the free-space conversion \(2\pi\,\delta L/\lambda\) and share the same phase-noise budget.

To first order in a static detuning, the phase-to-amplitude conversion of Appendix~\ref{app:detuning} is
\begin{equation}
	H_Y(\Omega)\simeq
	\frac{\ii\Omega\Delta}{(\kappa/2-\ii\Omega)^2},
\end{equation}
so that \(|H_Y|^2\rightarrow(\Delta/\Omega)^2\) for \(\Omega\gg\kappa\).  For \(\sigma_\Delta/\kappa=0.01\), the conversion factor at the equal-efficiency point is \(\sim4\times10^{-6}\).  The associated noise then depends on the laser frequency-noise spectrum.  The off-resonant LO quadrature transfer is summarized in Appendix~\ref{app:detuning}.

\paragraph{Spatial mode matching and detector efficiency.}
The cavity spatially filters the LO transmitted from M2, but it does not spatially filter the far-detuned signal sidebands, which are promptly reflected at M1.  The signal must therefore be matched independently to the same cavity eigenmode.  Its overlap with the transmitted LO, together with the photodiode efficiency, reduces the detected signal efficiency according to
\begin{equation}
	\eta_{\mathrm{sig}}
	=\eta_{\mathrm{pd}}\eta_{\mathrm{mm}}
	\eta_{\mathrm{cav}}.
\end{equation}
When the same \(\eta_{\mathrm{pd}}\) and \(\eta_{\mathrm{mm}}\) are assumed for both readouts, these factors must be included in a numerical comparison but do not alter the ideal relation in Eq.~\eqref{eq:equal-efficiency-filter}.

The single-mode analysis also presumes that no higher-order transverse mode is co-resonant within the analysis band.  Higher-order modes resonate at offsets set by the round-trip Gouy phase, and because the band spans an appreciable fraction of the free spectral range, the transverse-mode spacing and polarization splitting must be chosen such that low-order resonances avoid the analysis band, as in output-mode-cleaner design \cite{Fricke2012}.  At a co-resonance, mode-mismatched LO light would be resonantly transmitted rather than filtered, causing a narrow degradation of the LO-noise suppression. Signal components in the same higher-order mode would likewise produce narrow efficiency dips with sideband-asymmetric response.

\paragraph{Detected dc power and intracavity power.}
When the detected carrier is dominated by the LO, equal detected LO power gives the same mean optical power on the single photodiode in the cavity-assisted and asymmetric-beam-splitter readouts.  The cavity-assisted readout therefore does not intrinsically improve detector saturation or dc dynamic range.  It instead reduces the LO power required at the optical input port.  This reduction comes at the cost of resonant intracavity buildup.  In the design of Sec.~VII, \(P_{\LO,\mathrm{det}}=5\,\mathrm{mW}\) and \(T_1\simeq0.304\%\) correspond to a circulating LO power incident on M1 inside the cavity of approximately \(P_{\LO,\mathrm{det}}/T_1\simeq1.6\,\mathrm{W}\) in the idealized optical model.  Absorption and power-dependent detuning may therefore impose a separate practical limit.  In particular, absorption of the circulating power converts LO power fluctuations into detuning fluctuations through photothermal expansion.  The corresponding low-frequency requirement follows from Eq.~\eqref{eq:phase-noise-mixing}.

\paragraph{Backscatter and isolation.}
Mirror backscatter couples the circulating field into the counter-propagating cavity mode, which is degenerate, resonantly enhanced, and exits M1 along the reversed signal-injection path toward the quantum-signal source.  For a round-trip power fraction \(s^2\) coherently scattered into the counter-propagating mode, a simple coherent-buildup estimate gives a retro-injected power through M1 of
\begin{equation}
	P_{\mathrm{back}}
	\simeq
	s^2\left(\frac{\mathcal F}{\pi}\right)^2P_{\LO,\mathrm{det}}.
\end{equation}
For illustration, limiting this power to \(1\,\mathrm{nW}\) requires \(s^2\) to be of order \(10^{-12}\) per round trip.  The acceptable retro-injected power is source dependent and must be determined for the specific squeezed-light source.  If the corresponding backscatter level cannot be achieved, the signal path must be isolated, and the isolator insertion loss enters \(\eta_{\mathrm{sig}}\) directly.  Because an isolator transmission of \(0.98\)--\(0.99\) would consume the cavity's residual efficiency margin in Table~\ref{tab:squeezing-budget}, the efficiency advantage over the filtered beam-splitter readout survives only with an experimentally established, device-specific backscatter bound or comparably lossless isolation.

Backscattered light in the squeezed-light injection path of GEO~600 has been shown to couple to the optical parametric oscillator and degrade the measured quantum-noise reduction \cite{Bergamin2023}.  That observation establishes the practical relevance of retro-injection, but it does not test the ring-cavity geometry or the estimate above.

\paragraph{Balanced homodyne detection.}
Ideal balanced homodyne detection provides the reference benchmark and is not generally outperformed by the passive cavity-assisted readout.  In a practical balanced receiver, residual LO noise is determined by the complex, frequency-dependent mismatch of the two detector channels, often reported as a frequency-dependent common-mode rejection ratio \cite{Stefszky2012,Milovancev2022}.  A quantitative comparison with a particular receiver requires its measured complex transfer functions and electronic-noise spectra.  The quantitative comparison in this work is therefore restricted to single-photodiode readouts and makes no claim of lower total noise than BHD.

For orientation, a balanced receiver with a \(2.5\,\mathrm{GHz}\) bandwidth and a \(66\,\mathrm{dB}\) common-mode rejection ratio at \(2\,\mathrm{GHz}\) has been reported \cite{Wang2026}.  This demonstrates that state-of-the-art BHD can provide LO-noise rejection exceeding the maximum \(56.1\,\mathrm{dB}\) of the Sec.~VII cavity example, although a frequency-by-frequency comparison requires the measured complex response of the balanced receiver.  The motivation for the cavity-assisted readout is therefore not stronger LO-noise rejection than state-of-the-art BHD, but the absence of a differential channel-matching requirement in regimes where that matching becomes limiting, as in the balanced operation limited above approximately \(500\,\mathrm{MHz}\) in Ref.~\cite{Wilken2025}.

\paragraph{M3 and phase control.}
The small transmission of the nominally high-reflecting M3 is included in \(\kappa_\ell\).  Leakage through M3 could provide an auxiliary monitor for cavity locking.  Detection-phase control would additionally require a coherent signal component or a frequency-shifted coherent control field. Frequency-shifted coherent control fields are established for locking the phase of squeezed vacuum relative to an LO \cite{Chelkowski2007}, although their use in the present two-port geometry would require a separate implementation study. Such a control field could be co-injected with the signal at an offset within the reflected band but outside the chosen analysis window.  It would then be promptly reflected and beat with the transmitted LO at the control-field frequency.  Modulation sidebands used to lock the cavity to the LO could likewise be placed near the antiresonance, where their leakage to the detector port is suppressed by Eq.~\eqref{eq:periodic-lo-transfer}.  The LO field reflected at M2 carries the technical LO sidebands with a suppressed carrier and provides a further monitor port, analogous to optical ac coupling \cite{Kwee2008}.  These are implementation options rather than part of the central readout principle.

\section{Discussion}

The cavity replaces the fixed splitting ratio of an asymmetric combiner with a frequency-dependent transformation.  In the ideal resonant, impedance-matched case, Eq.~\eqref{eq:complementarity} shows that the signal efficiency and normalized technical LO-noise transfer sum to unity.  At equal signal efficiency, the cavity-assisted and asymmetric-beam-splitter readouts therefore have the same vacuum penalty.  The cavity-assisted readout instead benefits from efficient transfer of the resonant LO carrier and optical rejection of technical LO sidebands.  It does not reduce fundamental shot noise.

The cavity-assisted readout is a band-selective quadrature measurement with a single photodiode.  It is intended for frequencies at which a matched photodiode pair is unavailable or its differential response cannot be maintained.  For example, balanced operation in Ref.~\cite{Wilken2025} was limited above approximately \(500\,\mathrm{MHz}\), whereas a single-detector configuration operated to \(6.4\,\mathrm{GHz}\).  Relative to an unfiltered asymmetric beam splitter, the cavity-assisted readout adds optical filtering of technical LO noise and, for the 99:1 comparison, requires roughly one hundredth of the incident LO power.  Relative to a separately filtered beam-splitter readout, it performs the filtering and field combination in one resonator and removes the fixed splitting-ratio bound on signal efficiency.

The cavity also defines the spatial mode of the transmitted LO.  The resonantly transmitted LO is projected onto the cavity eigenmode, so imperfect LO injection primarily reduces detected LO power rather than the spatial purity of the transmitted component.  The promptly reflected signal sidebands are not spatially filtered and must still be matched to this mode.  Once this match is established, however, no additional signal--LO overlap is required at a separate beam splitter.  A conventional beam-splitter readout with a separate LO mode cleaner instead requires both LO matching to the cleaner and subsequent signal--LO overlap at the combining beam splitter.  The cavity-assisted layout therefore removes one downstream mode-matching step and one source of differential pointing and wavefront error.  Signal mismatch still acts as optical loss and can return residual signal-carrier power to the detector port.

Suppression of the resonant signal carrier gives the readout an optically ac-coupled character.  It is suited to quadrature measurements in a selected analysis-frequency band for which the signal mean field and near-carrier response are not observables of interest.  Electronic high-pass filtering cannot reproduce this optical rejection because it acts only after photodetection.  The signal carrier still contributes to the detected dc power, and beating between the carrier and optical noise sidebands has already generated photocurrent fluctuations before electronic filtering.

The practical example also exposes the band selectivity imposed by the periodic cavity resonances.  The linewidth and FSR must therefore be chosen for the target analysis band.  Higher finesse moves the reflected-sideband region toward lower frequencies and strengthens LO-noise rejection, but also increases intracavity buildup and sensitivity to detuning noise and fixed optical loss.  These constraints modify the readout band rather than the intrinsic bandwidth of the squeezed state.  Continuous coverage of an analysis band extending to \(6.4\,\mathrm{GHz}\) within a single inter-resonance window requires a correspondingly short resonator.  For example, \(\FSR\gtrsim6.4\,\mathrm{GHz}\) corresponds to \(L_{\mathrm{rt}}\lesssim47\,\mathrm{mm}\).  Alternatively, higher inter-resonance windows can be used through the periodic response of Appendix~\ref{app:detuning}.  Because the in-band reflection \(r(\Omega)\) retains a frequency-dependent phase, time-domain or temporal-mode reconstruction additionally requires calibration of the complex transfer function.  The present analysis concerns stationary quadrature spectra.

As a possible extension, cavity-assisted readout may be useful for displaced squeezed states.  For a coherent signal amplitude \(\alpha\), the linearized balanced-homodyne difference current contains an LO-quadrature fluctuation term proportional to \(\alpha\).  At exact impedance matching, \(r(0)\alpha=0\) at the cavity detector port, so this coupling is absent.  The coherent signal carrier then contributes neither to the detected dc power nor to this LO-noise coupling at linear order.  Accurate impedance and spatial mode matching are essential because any residual same-mode signal carrier grows in importance with \(\alpha\).  This extension is useful only when the displacement and low-frequency response need not be measured.  Phase control could use an auxiliary port or the frequency-shifted coherent control field discussed in Sec.~VIII.  A quantitative comparison lies beyond the present work.

A further extension is to inject the LO on an adjacent longitudinal resonance, converting the readout into a cavity-assisted heterodyne measurement, analogous to the classical two-port configuration of Ref.~\cite{Spector2024}.  Its analysis lies beyond the present scope.

The passive cavity-assisted readout cannot generally outperform ideal BHD. Its advantage is instead the removal of the differential detector-channel matching requirement while retaining high signal efficiency and providing optical LO-noise filtering.  Whether it outperforms a particular practical balanced receiver depends on that receiver's measured complex response and on the complete cavity-lock and loss budget.

\section{Conclusion}

This study shows that an impedance-matched traveling-wave cavity can combine a resonant LO carrier with off-resonant quantum-signal sidebands at a single photodiode, while their relative optical phase selects the measured quadrature. In the analysis band several cavity linewidths from the carrier and away from neighboring longitudinal resonances, the signal efficiency approaches unity while technical LO sidebands are optically suppressed.  At equal detected LO power and signal efficiency, the cavity-assisted readout requires substantially less incident LO power and provides stronger suppression of technical LO amplitude noise than a strongly asymmetric beam splitter with an unfiltered LO. Its signal efficiency also continues to approach unity rather than remaining fixed by a splitting ratio.  The resulting architecture provides a band-selective single-detector readout for high-frequency quadrature measurements without requiring matched photodetection channels.  These advantages come at the cost of resonant intracavity buildup and increased requirements on cavity loss, detuning stability, linewidth, and free spectral range.

\begin{acknowledgments}
I thank Nived Johny for helpful discussions and comments on the manuscript. This work was funded by the Carl-Zeiss-Stiftung (CZS Center QPhoton).
\end{acknowledgments}

\section*{Data Availability}

No data were created or analyzed in this study.  The numerical plots can be reproduced directly from the equations and parameters given in the manuscript.

\appendix

\section{Detuned and periodic cavity transfer}
\label{app:detuning}

For a static detuning \(\Delta\), normalize the upper and lower LO sidebands to the transmitted carrier:
\begin{align}
 T_+(\Omega)
 &=\frac{t(\Omega;\Delta)}{t(0;\Delta)},\\
 T_-^*(\Omega)
 &=\frac{t^*(-\Omega;\Delta)}{t^*(0;\Delta)}.
\end{align}
The detected amplitude-quadrature response to LO amplitude and phase modulation is
\begin{align}
 H_X(\Omega)
 &=\frac{T_+(\Omega)+T_-^*(\Omega)}{2},\\
 H_Y(\Omega)
 &=\frac{T_+(\Omega)-T_-^*(\Omega)}{2\ii}.
\label{eq:detuned-lo-transfer}
\end{align}
At \(\Delta=0\), the two sidebands are conjugate symmetric, \(H_Y=0\), and \(H_X=t(\Omega)/t(0)\), which recovers Eq.~\eqref{eq:lo-filter}.  For nonzero detuning, \(H_Y\neq0\), so LO phase noise can be converted into detected amplitude noise.  A constant detuning defines a calibratable transfer function.  Fluctuations of that detuning produce additional noise.

For frequencies that are not small compared with the FSR, let \(\tau_{\mathrm{rt}}=1/\FSR\) and let \(\rho\) denote the resonant round-trip amplitude factor.  The exact normalized LO power transfer on carrier resonance is
\begin{equation}
 \left|H_{\LO}^{\mathrm{per}}(\Omega)\right|^2
 =\frac{(1-\rho)^2}
 {1+\rho^2-2\rho\cos(\Omega\tau_{\mathrm{rt}})}.
\label{eq:periodic-lo-transfer}
\end{equation}
For equal couplers with power transmission \(T\) and round-trip internal power loss \(\mathcal L_{\mathrm{int}}\), define \(r_c=\sqrt{1-T}\), \(a_\ell=\sqrt{1-\mathcal L_{\mathrm{int}}}\), and \(\rho=r_c^2a_\ell\).  The exact Airy finesse is related to \(\rho\) by
\begin{equation}
 \sin\!\left(\frac{\pi}{2\mathcal F}\right)
 =\frac{1-\rho}{2\sqrt{\rho}}.
\label{eq:finesse-rho}
\end{equation}
The corresponding signal-port amplitude reflection is
\begin{equation}
 r_{\mathrm{eq}}^{\mathrm{per}}(\Omega)
 =r_c\frac{1-a_\ell e^{\ii\Omega\tau_{\mathrm{rt}}}}
 {1-\rho e^{\ii\Omega\tau_{\mathrm{rt}}}}.
\label{eq:periodic-equal-coupler-reflection}
\end{equation}
This is the signal response used for the practical design in Fig.~\ref{fig:representative-transfer}. For the same periodic model,
\begin{align}
 |t^{\mathrm{per}}(\Omega)|^2
 &=T_{\LO}(0)\left|H_{\LO}^{\mathrm{per}}(\Omega)\right|^2,\notag\\
 |l^{\mathrm{per}}(\Omega)|^2
 &=1-|r_{\mathrm{eq}}^{\mathrm{per}}(\Omega)|^2
  -|t^{\mathrm{per}}(\Omega)|^2 .
\label{eq:periodic-loss-coupling}
\end{align}
The effective loss-vacuum coefficient in Eq.~\eqref{eq:periodic-loss-coupling} restores the exact periodic output commutator. For an exactly impedance-matched cavity, distinct from the slightly undercoupled equal-coupler example above, the periodic responses satisfy \(\eta_{\mathrm{cav}}^{\mathrm{per}}=1-|H_{\LO}^{\mathrm{per}}|^2\). For \(|\Omega\tau_{\mathrm{rt}}|\ll1\) and \(1-\rho\simeq\kappa\tau_{\mathrm{rt}}/2\), Eq.~\eqref{eq:periodic-lo-transfer} reduces to Eq.~\eqref{eq:lo-filter}. At antiresonance, its high-finesse limit is \(|H_{\LO}^{\mathrm{per}}|^2\simeq[\pi/(2\mathcal F)]^2\).

\bibliography{references}

\end{document}